\documentclass[10pt,twocolumn,english,aps,pra,showpacs,superscriptaddress,groupaddress,floatfix,graphics,graphicx,longbibliography]{revtex4-1}

\usepackage[T1]{fontenc}
\usepackage{color}
\usepackage[english]{babel}
\usepackage{float}
\usepackage{amsmath}
\usepackage{amssymb}
\usepackage{graphicx}
\usepackage{xcolor}
\usepackage[dvipsnames]{xcolor}
\usepackage[bookmarks]{hyperref}
\hypersetup{colorlinks,linkcolor=blue,citecolor=blue,urlcolor=blue}
\usepackage[all]{hypcap}
\usepackage{orcidlink}

\begin{document}	

\title{Proximity-Induced Spin-Orbit Torque in Graphene on a Trigonal CrSBr Monolayer}

%%%%%%%%%%%%%%%%%%%%%%%%%%%%%%%%%%%%%%%%%%%%%%%%%%%%%%%%%%%%%%%%%%%%%%%%%%%%%%%%
\author{Maedeh Rassekh,\orcidlink{0000-0003-1093-7502}}
\email[Corresponding author: ]{maedeh.rassekh@upjs.sk}
\affiliation{Institute of Physics, Pavol Jozef Šafárik University in Košice, Park Angelinum 9, 04001 Košice, Slovakia}

\author{Martin Gmitra,\orcidlink{0000-0003-1118-3028}}
\affiliation{Institute of Physics, Pavol Jozef Šafárik University in Košice, Park Angelinum 9, 04001 Košice, Slovakia}
\affiliation{Institute of Experimental Physics, Slovak Academy of Sciences, 04001 Košice, Slovakia}

%%%%%%%%%%%%%%%%%%%%%%%%%%%%%%%%%%%%%%%%%%%%%%%%%%%%%%%%%%%%%%%%%%%%%%%%%%%%%%%%
\begin{abstract}
We present a first-principles and quantum transport study of proximity-induced spin-orbit torque (SOT) in graphene on a trigonal CrSBr monolayer. Density functional theory combined with nonequilibrium Green’s function calculations shows that the CrSBr substrate induces spin polarization and a sizable exchange splitting in the graphene Dirac states. The resulting current-driven spin density in graphene generates a self-SOT on the Dirac electrons. The proximity-induced exchange field breaks time-reversal symmetry and gives rise to a purely odd SOT component, while the even contribution vanishes. The torque magnitude exhibits a strong angular dependence with phase shifts arising from the noncollinearity between the CrSBr magnetization and the induced magnetic moments in graphene. Monte Carlo simulations based on the calculated exchange parameters predict a Curie temperature of 
$T_{\rm c} \approx 304\ \mathrm{K}$, confirming the robustness of ferromagnetism in the trigonal CrSBr monolayer. These results identify graphene/CrSBr heterostructures as a promising platform for room-temperature two-dimensional spintronics.
\end{abstract}

\maketitle
%%%%%%%%%%%%%%%%%%%%%%%%%%%%%%%%%%%%%%%%%%%%%%%%%%%%%%%%%%%%%%%%%%%%%%%%%%%%%%%%
\section{Introduction}

Spin-orbit torques (SOTs) have emerged as a central mechanism for controlling magnetization in non-volatile spintronic devices \cite{miron2011perpendicular,liu2012spin,garello2013symmetry}. While conventional SOT-MRAM relies on heavy metals to generate spin currents, achieving efficient and tunable torques at reduced energy cost requires exploring alternative platforms. Two-dimensional (2D) materials and their van der Waals heterostructures provide such opportunities, not only because of their high mobility and long spin coherence, but also due to their ability to host proximity-induced spin-orbit and magnetic effects \cite{han2014graphene,safeer2019room,benitez2020tunable}.

Building on this, stacking 2D materials in van der Waals heterostructures has opened a versatile platform for realizing all-2D memory and logic devices \cite{khokhriakov2020two,khokhriakov2022multifunctional} and for exploring novel phenomena such as charge-spin interconversion (CSI) \cite{jafari2022spin,wang2016giant}. Efficient and controllable CSI is essential for spintronics applications, as it directly underpins SOT generation. The microscopic origins of CSI include the Rashba-Edelstein effect \cite{edelstein1990spin} and the Spin Hall effect \cite{sinova2015spin,hirsch1999spin}, both of which create non-equilibrium spin accumulation that can drive current-induced magnetization switching.

Within this broader class of 2D heterostructures, graphene stands out as an exceptional channel material due to its ultrahigh mobility and long spin-coherence lengths \cite{tombros2007electronic,han2014graphene}. A central challenge, however, is understanding how proximity effects modify graphene’s electronic and spin properties. Considerable efforts have therefore focused on engineering graphene via coupling to other 2D materials \cite{safeer2019room,benitez2020tunable}. For instance, interfacing with transition metal dichalcogenides enhances spin-orbit coupling and enables efficient CSI \cite{gmitra2016trivial,gmitra2017proximity}, whereas coupling to 2D magnetic semiconductors induces magnetic proximity effects, leading to spin polarization and exchange splitting of the Dirac states \cite{zhong2017van,wang2015proximity}.

Among the emerging two-dimensional magnetic semiconductors, CrSBr stands out as a highly promising material. According to the C2DB database \cite{gjerding2021recent,haastrup2018computational}, its monolayer can exist in two dynamically stable phases. The orthorhombic \textit{Pmmn} phase is well established and has been extensively studied both experimentally \cite{lee2021magnetic,rizzo2022visualizing} and theoretically  \cite{yang2022first,yang2021triaxial}; it exhibits a layered structure \cite{yang2022first}, a sizable band gap \cite{yang2021triaxial}, and robust ferromagnetism with a Curie temperature of about 146~K \cite{lee2021magnetic}.
The trigonal (\textit{P3m1}) monolayer has not been experimentally explored.

In this work, we theoretically examine the magnetic properties of the trigonal CrSBr monolayer and predict its Curie temperature. We further integrate it with graphene, forming a heterostructure that offers a compelling platform for proximity-induced spin-dependent properties and current-driven torques.

Using a combination of density functional theory (DFT), atomistic spin simulations, and non-equilibrium Green's function (NEGF) transport calculations, we systematically explore the interplay between electronic structure and torque generation. We find that the trigonal CrSBr monolayer exhibits robust ferromagnetism near room temperature and induces a sizable exchange field in graphene, leading to spin splitting of the Dirac states. 
Importantly, the resulting spin-orbit torques are dominated by the odd component, with angular dependencies that exhibit phase shifts as a result of an interplay between general magnetization direction in CrSBr and proximity-induced magnetic moments in graphene.

%%%%%%%%%%%%%%%%%%%%%%%%%%%%%%%%%%%%%%%%%
\section{Methods}\label{methods}
%---------------------------------------------------------
\begin{figure*}[t]
    \centering
    \includegraphics[width=0.8\textwidth]{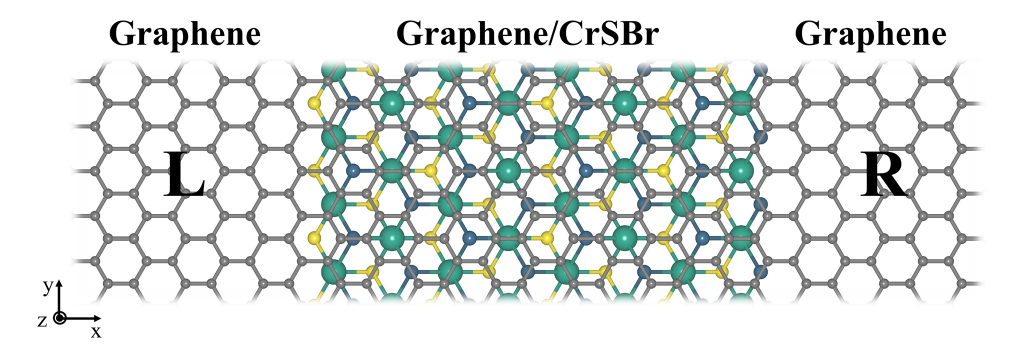}
    \caption{A Schematic view of the device composed of a central region made of graphene/CrSBr heterostructure attached to the left (L) and right (R) graphene leads for conducting unpolarized charge current into the graphene
within the central region.}
    \label{fig:Device}
\end{figure*}
%---------------------------------------------------------
\subsection{Computational details of DFT calculations}
We performed density functional theory (DFT) calculations \cite{hohenberg1964inhomogeneous,kohn1965self}, including spin-orbit coupling using the OpenMX code \cite{ozaki2003variationally,ozaki2004numerical,ozaki2005efficient,lejaeghere2016reproducibility}, which is based on norm-conserving pseudopotentials with partial core corrections \cite{bachelet1982pseudopotentials,troullier1991efficient,kleinman1982efficacious,blochl1990generalized,morrison1993nonlocal,vanderbilt1990soft}. The basis functions are constructed from linear combinations of pseudo-atomic orbitals (LCPAO) \cite{ozaki2004numerical}. The basis sets used in this work were specified as S7.0-s2p2d1f1 for sulfur, Cr6.0-s3p2d1 for chromium, and Br7.0-s3p2d2 for bromine. These notations indicate the cutoff radius (in Bohr) and the number of primitive orbitals used for each angular momentum channel in the LCPAO basis set. For instance, Cr6.0-s3p2d1 means that the cutoff radius is 6.0 Bohr, and the basis orbitals for the chromium atom include three primitive functions for the $s$ orbitals, two for $p$, and one for $d$ orbital.
We used the generalized gradient approximation (GGA) with the Perdew-Burke-Ernzerhof (PBE) exchange-correlation functional \cite{perdew1996generalized}. All calculations were performed until the change in total energy between successive iterations was smaller than $ 10^{-6} $ Hartree.
To prevent interactions between periodic images in the out-of-plane direction, a vacuum spacing of 30 Å was applied along the $z$-axis.

% -------------------------------------------------------------------
\subsection{Quantum transport calculations of the torque}
In quantum mechanics, the expectation value of any physical quantity, such as spin or torque, can be calculated using the trace of the product of the density matrix and the corresponding operator:
\begin{equation}
	\mathbf{O} = \mathrm{Tr}[\rho\,\hat{\mathbf{O}}]
\end{equation}
In steady-state out-of-equilibrium conditions, the nonequilibrium density matrix can be obtained from the lesser Green’s function $G^<(E)$ within the Keldysh formalism~\cite{dolui2020proximity}:
\begin{equation}
	\rho_{\rm neq} = \frac{1}{2\pi i} \int_{-\infty}^{\infty} G^<(E) \, dE
\end{equation}
where
\begin{equation}
	G^<(E) = i G^{\rm r}(E) \left[f_{\rm L}(E)\Gamma_{\rm L}(E) + f_{\rm R}(E)\Gamma_{\rm R}(E)\right] G^{\rm a}(E)
\end{equation}
Here, $G^{\rm r}(E)$ and $G^{\rm a}(E)$ are the retarded and advanced Green's functions, which are defined, respectively, as:
\begin{align}
	G^{\rm r}(E) &= \left[(E + i\eta) - H - \Sigma_{\rm L}^{\rm r}(E, V_{\rm L}) - \Sigma_{\rm R}^{\rm r}(E, V_{\rm R})\right]^{-1} \\
	G^{\rm a}(E) &= \left[(E - i\eta) - H - \Sigma_{\rm L}^{\rm a}(E, V_{\rm L}) - \Sigma_{\rm R}^{\rm a}(E, V_{\rm R})\right]^{-1}
\end{align}
where $\eta$ is a small positive infinitesimal ensuring causality, and $\Sigma_{\rm L,R}^{\rm r,a}$ are the retarded and advanced self-energy matrices describing the coupling to the left and right leads, see Fig.~\ref{fig:Device} for the considered two-terminal setup with current along the $x$-direction. In non-orthogonal basis sets, the terms $(E \pm i\eta)$ must be replaced by $(E \pm i\eta)S$, with $S$ denoting the overlap matrix.

In the \textbf{elastic transport regime}, where dephasing processes are negligible, the lesser Green’s function is expressed solely in terms of the retarded Green’s function, with $ G^{\rm a}(E) = [G^{\rm r}(E)]^\dagger $.
The coupling matrices are defined as:
\begin{equation}
	\Gamma_{\rm L,R} = i\left[\Sigma_{\rm L,R}^{\rm r} - \Sigma_{\rm L,R}^{\rm a}\right].
\end{equation}
Assuming the bias voltage, $ V_{\rm b} = V_{\rm L} - V_{\rm R} $, induces a rigid shift of the electronic structure in the leads, the self-energies become~\cite{rassekh2021charge,nikolic2020first}:
\begin{align}
	\Sigma_{\rm L}(E, V_b) &= \Sigma_{\rm L}(E - eV_{\rm L}) \\
	\Sigma_{\rm R}(E, V_b) &= \Sigma_{\rm R}(E - eV_{\rm R}).
\end{align}
The functions $f_{\rm L,R}(E)=f(E-\mu_{L,R})$ represent the Fermi-Dirac distributions of the left and right leads. Under an applied \textbf{symmetric bias} \(V_b\), the chemical potentials shift as $\mu_L = \mu + eV_b/2$, and $\mu_R = \mu - eV_b/2$.

In the \textbf{linear-response regime} $(eV_b \ll E_F)$, the Fermi-Dirac distribution functions can be expanded as:
\begin{align*}
	f_L(E) \approx f(E-eV_b/2) \approx f(E)-\frac{\partial f}{\partial E} \frac{eV_b}{2} + \dotsm\\
	f_R(E) \approx f(E+eV_b/2) \approx f(E)+\frac{\partial f}{\partial E} \frac{eV_b}{2} + \dotsm
\end{align*}
This allows us to decompose the density matrix into the Fermi sea and surface contributions~\cite{mahfouzi2012how, dolui2020spin}:
\begin{align}
\rho_{\rm sea} &= \frac{i}{2\pi} \int f(E)\big(G(E)-G^\dagger(E)\big) dE \\
\rho_{\rm surf} &= \frac{eV}{4\pi} \int \left(-\frac{\partial f}{\partial E} \right) G(E)\big[ \Gamma_{\rm L}(E)-\Gamma_{\rm R}(E) \big] G^\dagger(E) dE
\end{align}
where $f(E)$ is the equilibrium Fermi-Dirac distribution~\cite{belashchenko2019first,ghosh2023perspective}. 
Here we note that this separation is not gauge invariant and was derived under the assumption of a symmetric bias voltage. As emphasized in Refs.~\cite{mahfouzi2012how,belashchenko2019first}, such decompositions into Fermi-sea and Fermi-surface contributions involve a specific choice of gauge, which is not unique but rather a convenient mathematical partitioning.

At \textbf{zero temperature}, $-\partial f / \partial E$ becomes a Dirac delta function at Fermi energy, $E = E_F$, simplifying the surface term to:
\begin{equation}
	\rho_{\rm surf} = \frac{eV_b}{4\pi} G(E_F)(\Gamma_L(E_F)-\Gamma_R(E_F))G^\dagger(E_F).
\end{equation}
Hence, the total nonequilibrium density matrix is:
\begin{equation}
	\rho_{\rm neq} = \rho_{\rm sea} + \rho_{\rm surface}.
\end{equation}
We note that to properly implement the above approach in the linear-response regime, one needs to include the voltage drop across the center region for the Fermi-sea term~\cite{joao2024reconciling}:
\[
G^{r}_{V_b} = \left[E - H - eU_i - \Sigma_L^r(E-eV_L)- \Sigma_R^r(E-eV_R)\right]^{-1}
\]
This term must be present even if the central active region is disorder-free.

The current-driven part of the nonequilibrium density matrix is defined as~\cite{nikolic2020first}
\begin{equation}
	\rho_{\rm CD} = \rho_{\rm neq} - \rho_{\rm eq},
\end{equation}
where
\begin{equation}
	\rho_{\rm eq} = -\frac{1}{2\pi i} \int \left[G_0^r(E) - G_0^r(E)^\dagger\right] f(E) dE
\end{equation}
and $G_0^r (E)$ is the retarded Green's function at zero bias voltage.
From $\hat{\rho}_{\mathrm{CD}}(k_y)$ one can calculate the current-driven spin density values as
\begin{equation}
{\mathbf{S}_{\mathrm{CD}}}(k_y) = 
\mathrm{Tr}\!\left[ \hat{\rho}_{\mathrm{CD}}(k_y)\, \hat{\boldsymbol{\sigma}}\, S^{-1} \right],
\label{eq:S_CD}
\end{equation}
where $S$ is the overlap matrix, $\hat{\boldsymbol{\sigma}}=(\hat{\sigma}_x,\hat{\sigma}_y,\hat{\sigma}_z)$ 
are the Pauli matrices.
The total spin-orbit torque vector is then obtained by integrating the contributions 
from all transverse momenta over the one-dimensional Brillouin zone (BZ):
\begin{equation}
\mathbf{T}_{\rm CD} = \frac{1}{\Omega_{\rm BZ}} 
\int_{\rm BZ} dk_y \,\Big[ \mathbf{S}_{\mathrm{CD}}(k_y) \times \mathbf{B}_{\rm XC}(k_y)\Big],
\label{eq:T_CD}
\end{equation}
where $\Omega_{\rm BZ}$ is the BZ length along $k_y$, and for a self-consistently converged ncDFT Hamiltonian represented in the LCAO basis, the exchange field matrices can be extracted from the spinor blocks as
\begin{align}
B_{\rm XC}^x &= \tfrac{1}{2}(H_{\uparrow\downarrow}+H_{\downarrow\uparrow}), \\
B_{\rm XC}^y &= \tfrac{1}{2i}(H_{\downarrow\uparrow}-H_{\uparrow\downarrow}), \\
B_{\rm XC}^z &= \tfrac{1}{2}(H_{\uparrow\uparrow}-H_{\downarrow\downarrow}).
\end{align}

%%%%%%%%%%%%%%%%%%%%%%%%%%%%%%%%%%%%%%%%%%%%%%%%%%%%%%%%%%%%%%%%%%%%%%%%%%%%%%%%
\section{Results and discussion}\label{results}
%---------------------------------------
\begin{figure}[b]
     \includegraphics[width=1\columnwidth]{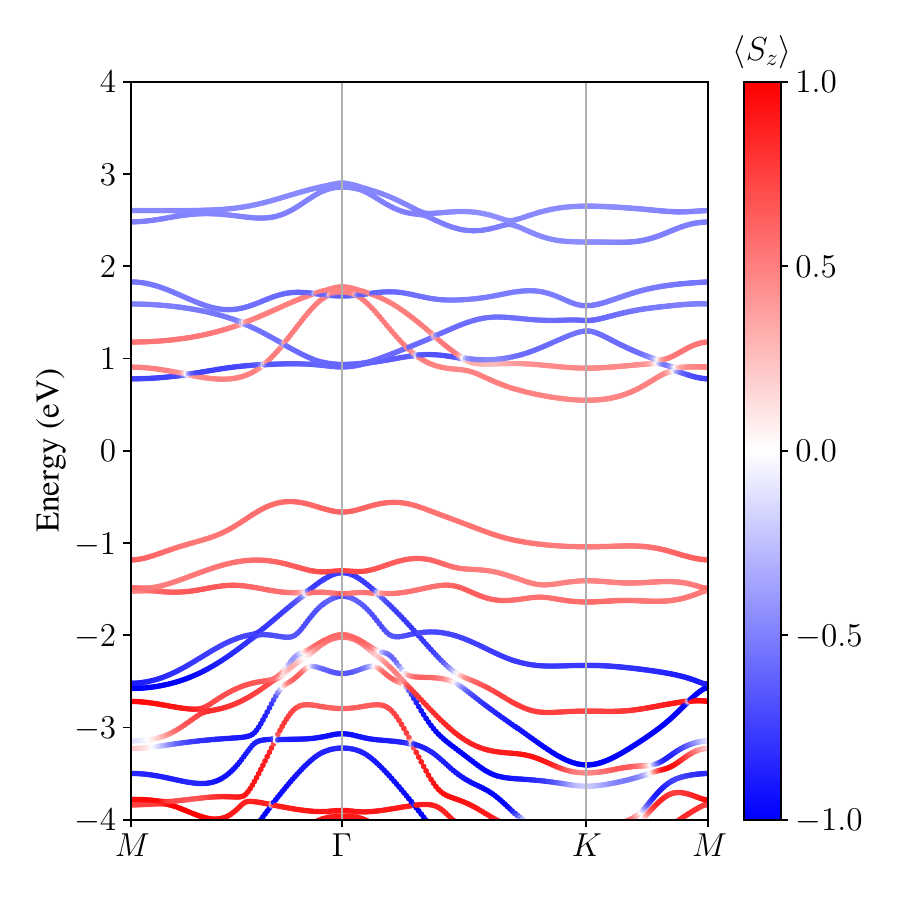}
    \caption{Calculated spin-resolved band structure of CrSBr monolayer unit cell.
    The color scale corresponds to $\langle S_z\rangle$ spin expectation value. The Fermi level is set to zero.}
    \label{fig:Band-CrSBr}
\end{figure}
%---------------------------------------
\subsection{Magnetic properties of trigonal CrSBr monolayer}
For the trigonal CrSBr monolayer in p3m1 layer group (\#69), we used lattice parameters as provided by the C2DB database \cite{gjerding2021recent,haastrup2018computational}, consisting of one chromium, one sulfur, and one bromine atom per unit cell, distinct from the more commonly known orthorhombic Pmmn structure.

Spin-resolved band structure for bare trigonal CrSBr monolayer is plotted in Fig.~\ref{fig:Band-CrSBr}. Our calculations show that the trigonal CrSBr monolayer is a ferromagnetic semiconductor with an indirect band gap of about 1.1 eV. The magnetic moment is concentrated mostly at the Cr atoms, which is about $3~\mu_B$.

% -------------------------------------------------------------------
Since the functionality of two-dimensional magnets depends not only on their ground-state properties but also on their thermal stability, we proceed to determine the Curie temperature.
To determine the Curie temperature \(T_{\rm c}\) of the trigonal CrSBr monolayer, we performed Monte Carlo simulations using the \textsc{VAMPIRE} software package \cite{evans2014atomistic,alzate2019optimal}, considering a rectangular supercell with length $L$ and exchange parameters obtained from \textsc{TB2J} \cite{he2021tb2j} calculations based on fully relativistic density functional theory (DFT). The isotropic exchange couplings \(J_{ij}\) were found to be strongly ferromagnetic for the nearest-neighbour Cr--Cr pairs (\(J_{\mathrm{iso}} \approx 13\)~meV).

%------------------------------------------
\begin{figure}[t]
    \centering
    \includegraphics[width=1\columnwidth]{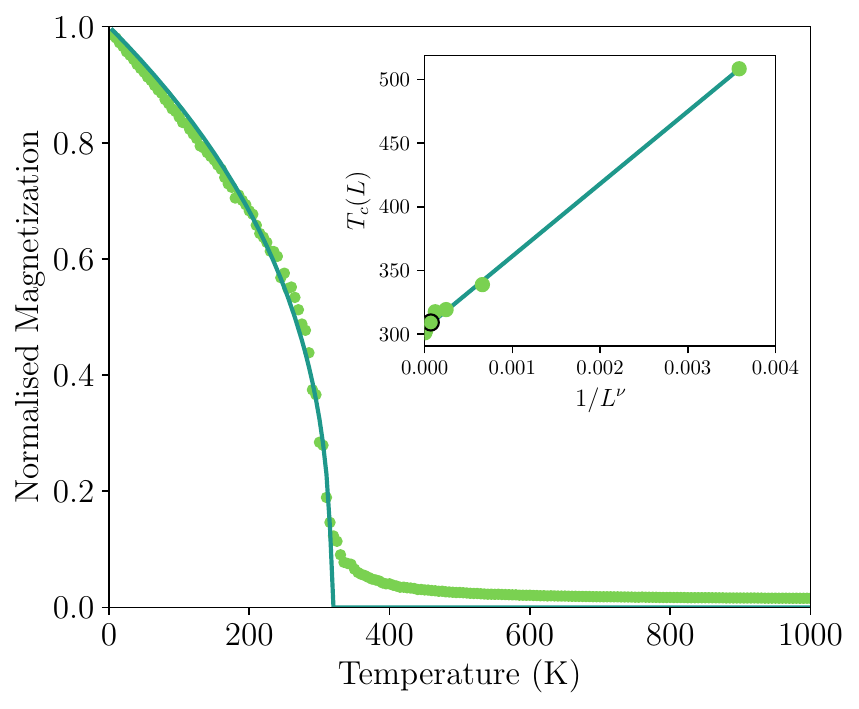}
    \caption{Temperature dependence of CrSBr monolayer magnetization obtained by Monte Carlo simulation (circles) and analytical fit
    (solid line) for a finite system with the linear size $L=50$.
    The fitted Curie temperature $T_{\rm c}(L=50) \approx 316.8\ \mathrm{K}$.
    The inset shows the finite-size scaling analysis of $T_{\rm c}(L)$ versus $1/L^{\nu}$ with $\nu \simeq 0.41$, yielding the critical temperature $T_{\rm c} \approx 304.4~\mathrm{K}$. The data point for $L=50$ is emphasised with a black circle.}
    \label{fig:curie-temperature}
\end{figure}
%------------------------------------------
The magnetization \(m(T)\) as a function of temperature was obtained by gradually increasing the temperature from 0~K to 1000~K in increments of 5~K, with sufficient equilibration and averaging steps at each point to ensure convergence. The Curie temperature for a given system size was first extracted by fitting the ferromagnetic region to the critical scaling relation for the absolute-valued mean magnetization
\begin{equation}
    \langle |m(T,L)|\rangle = \left( 1 - \frac{T}{T_{\rm c}(L)} \right)^\beta ,
    \label{eq:analytic-m(T)}
\end{equation}
with pseudocritical temperature $T_{\rm c}(L)$ and the critical exponent $\beta$ treated as fitting parameters. In Fig.~\ref{fig:curie-temperature} we show the temperature dependence of the normalized $\langle |m(T,L)|\rangle$ for $L=50$. This approach yielded pseudocritical temperature $T_{\rm c} \approx 316.8\ \mathrm{K}$ and $\beta \approx 0.34$.
Employing finite-size scaling analysis, we determine critical temperature $T_{\rm c}$ for $L\to\infty$ by fitting the pseudocritical temperatures to $T_{\rm c}(L) = T_{\rm c} + a L^{-1/\nu}$, where $a$ is a non-universal constant, and $\nu$ is the critical exponent of the correlation length. A nonlinear fit yielded $T_{\rm c} = 304.4\ \mathrm{K}$, 
%$a = 5.68\times 10^4$, 
and $\nu \simeq 0.41$,
see inset of Fig.~\ref{fig:curie-temperature}.
We note that the estimated Curie temperature of CrSBr monolayer is remarkably close to room temperature, and significantly higher than the Curie temperatures reported for prototypical van der Waals magnets such as CrI\(_3\) (\(T_{\rm c} \sim 45\ \mathrm{K}\)) \cite{huang2017layer,rassekh2020remarkably} and Cr\(_2\)Ge\(_2\)Te\(_6\) (\(T_{\rm c} \sim 61\ \mathrm{K}\)) \cite{gong2017discovery}.

To further validate our methodology, we applied the same computational
workflow to the experimentally realized \textit{Pmmn} CrSBr monolayer, obtaining $T_{\mathrm{c}} \approx 183$~K. 
While this value exceeds the experimental $T_{\mathrm{c}} \approx 146$~K \cite{lee2021magnetic},
it is in good agreement with previous theoretical predictions \cite{yang2021triaxial}.

% -------------------------------------------------------------------
\subsection{Electronic structure of graphene/CrSBr heterostructure}

\begin{figure}[b]
     \includegraphics[width=1\columnwidth]{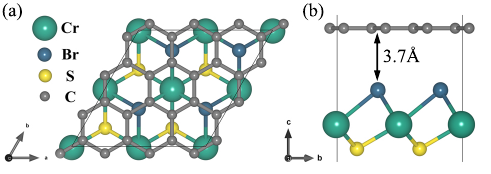}
    \caption{Optimized graphene/CrSBr heterostructure: (a) top view and (b) side view. The $3 \times 3$ graphene supercell matched with a $2 \times 2$ CrSBr supercell, and a uniaxial strain of approximately 3\% was applied to the graphene layer to achieve lattice commensurability. The structure was fully relaxed, resulting in an interlayer distance of 3.7~\AA.}
    \label{fig:CrSBr-Gr}
\end{figure}

To construct the commensurate heterostructure, a $3\times 3$ supercell of graphene was matched with a $2\times 2$ supercell of trigonal CrSBr. Due to the lattice mismatch between the two materials, a uniaxial strain of approximately $3 \%$ was applied to the graphene layer to ensure proper alignment, see Fig.~\ref{fig:CrSBr-Gr}. The structure was then fully relaxed, and the interlayer distance between graphene and CrSBr converged to 3.7~\AA~after structural optimization.
%--------------------------------------
%--------------------------------------

\begin{figure}[t]
     \includegraphics[width=1\columnwidth]{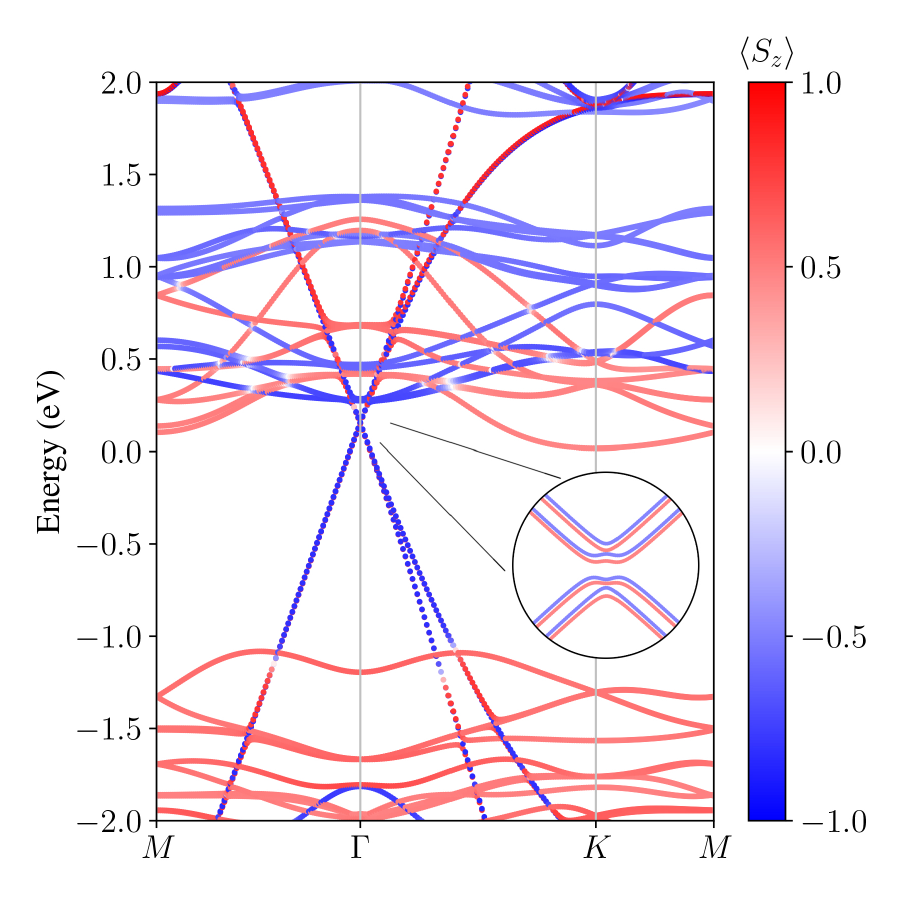}
    \caption{Calculated spin-resolved band structure of graphene/CrSBr heterostructure. Zoom shows band spin splitting topology of graphene states near the Dirac point. The color scale corresponds to $\langle S_z\rangle$ spin expectation value. The Fermi level is set to zero.}
    \label{fig:Band-CrSBr-Gr}
\end{figure}

The spin-resolved band structure of the graphene/CrSBr heterostructure is shown in Fig.~\ref{fig:Band-CrSBr-Gr}. The linear Dirac-like bands originating from graphene appear near the Fermi level. Due to the use of a $3 \times 3$ graphene supercell in the heterostructure construction, the Brillouin zone is folded, and the Dirac cone is consequently relocated from the $K$-point to the $\Gamma$-point.

%---------------------------------------
%---------------------------------------

Notably, the Dirac bands exhibit a clear spin splitting, with distinct spin-up (red) and spin-down (blue) components. This splitting is characteristic of the exchange interaction caused by the underlying CrSBr layer, which breaks time-reversal symmetry and lifts the spin degeneracy of the graphene states. Additionally, the Dirac point is visibly shifted above the Fermi level, indicating that the graphene layer experiences n-type doping due to charge transfer from the CrSBr substrate.

In general, proximity to materials with strong spin-orbit coupling or broken sublattice symmetry can lead to gap opening or spin splitting in graphene through several proximity mechanisms, such as interface-induced SOC~\cite{zollner2024proximity,gmitra2017proximity}, exchange fields~\cite{zollner2016theory,jafari2025exchange}, and %Kekulé 
distortions~\cite{lin2017competing}.
In the graphene/CrSBr heterostructure, the presence of graphene does not modify the Curie temperature of the CrSBr.
The calculated total charge transfer from graphene to CrSBr is $\Delta Q = 2.288\times 10^{-3}\,e/$\AA$^{2}$, corresponding to a carrier density of $n \approx 2.3\times 10^{13}\,\mathrm{cm^{-2}}$, which is smaller than doping levels typically required to substantially modify exchange parameters and Curie temperature in two-dimensional magnets \cite{Xie2023:PCCP}.
The analysis based on Mulliken population shows that the transferred charge is almost entirely accommodated on the interfacial Br atoms.
Each Br atom gains approximately $0.025\,e$, while the changes on S atoms and, importantly, on the magnetic Cr atoms remain below $10^{-3}\,e$. 
This demonstrates that graphene-induced doping primarily affects the outer Br layer, leaving the magnetic moments on Cr atoms unaffected.

% -------------------------------------------------------------------
\subsection{Proximity-induced SOT in graphene/CrSBr}

\begin{figure}[b]
    \centering
    \includegraphics[width=1\columnwidth]{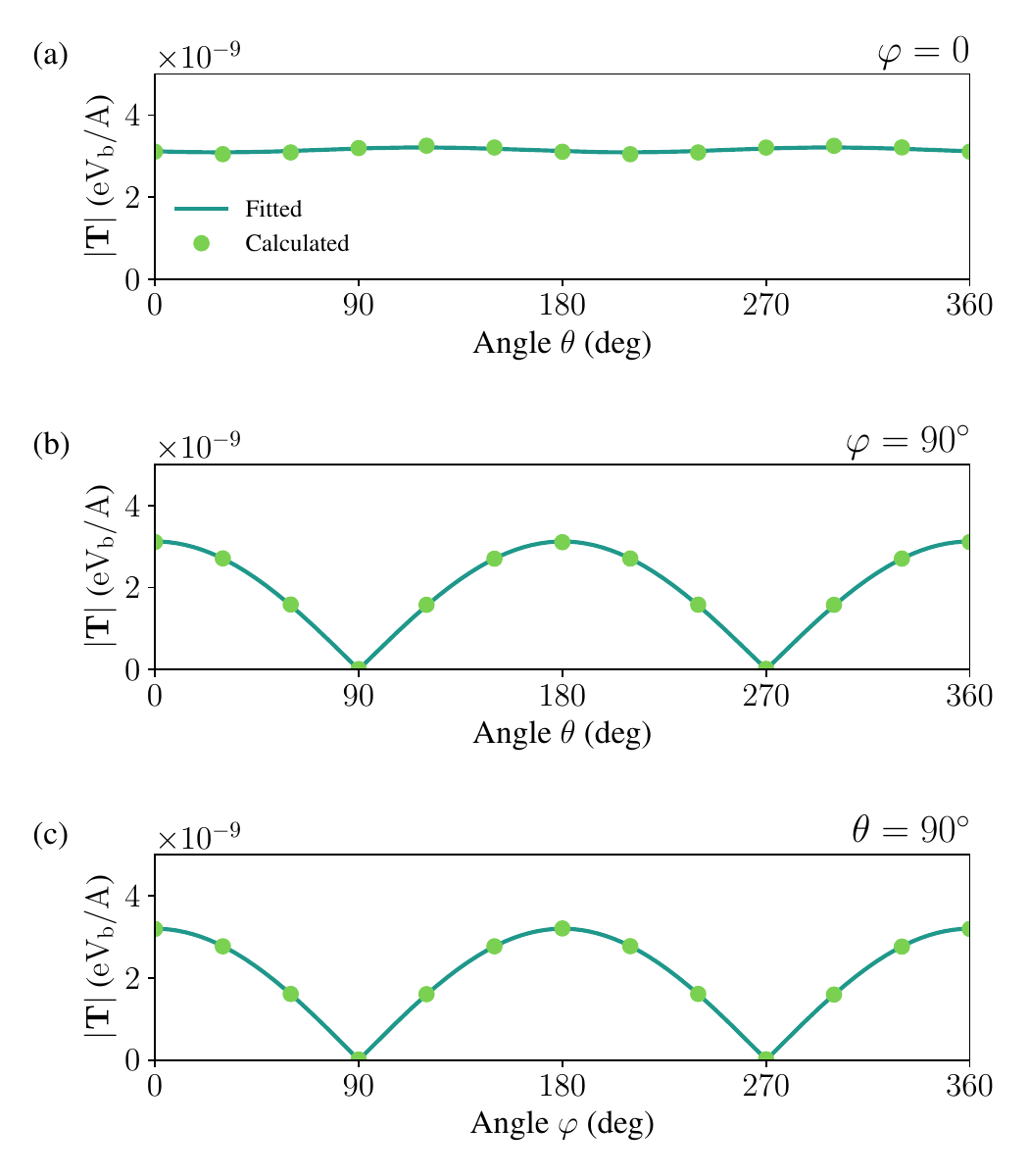}
    \caption{Azimuthal ($\theta$) and polar ($\varphi$) angle dependence of the magnitude of the calculated odd spin-orbit torque component, $|\mathbf{T}|$, expressed in units of $eV_b/A$ at $E_F = 0.35 $ eV. Here, $A$ denotes the area of a single hexagon in the graphene lattice. The magnetization direction is given by $\mathbf{m} = (\sin\theta \cos\varphi, \sin\theta \sin\varphi, \cos\theta)$ on the CrSBr layer. The rotation of $\mathbf{m}$ is restricted to (a) the $xz$ plane, (b) the $yz$ plane, and (c) the $xy$ plane. Solid green lines are fit to NEGF+ncDFT-computed SOT values (dots) using Eq.~\ref{eq:series} with fitting paramers from Table \ref{tab:coeffs}.}
    \label{fig:SOT}
\end{figure}

To calculate the current-driven spin density and spin-orbit torque (SOT) in the two-terminal device as shown in Fig.~\ref{fig:Device}, where a finite central region of the graphene/CrSBr heterostructure is attached to semi-infinite graphene leads, we employ a first-principles quantum transport framework, which combines nonequilibrium Green functions (NEGFs) \cite{rassekh2021charge} with noncollinear density functional theory (ncDFT) calculations \cite{macdonald1979relativistic,bachelet1982pseudopotentials,theurich2001self}.  A small symmetric bias $eV_b = 2$~meV is applied across the leads.

The total spin-orbit torque can be decomposed into components that are even ($\mathbf{T}^{e}$) and odd ($\mathbf{T}^{o}$) with respect to the magnetization direction $\mathbf{m}$, i.~e, $ \mathbf{T} = \mathbf{T}^{e} + \mathbf{T}^{o} $.
These contributions are obtained directly from Eq.~(\ref{eq:T_CD}) by using the respective components of the density matrix \cite{dolui2020proximity}.
Because the system lacks a conducting bulk, no vertical spin Hall current along the $z$ axis could generate an even torque contribution. 
Furthermore, since we assume ballistic transport, interfacial mechanisms requiring backscattering of electrons are absent. Therefore, we find that the even torque $\mathbf{T}^e \rightarrow 0$, and the total SOT is dominated by the odd component.
The magnitudes of the odd SOT components, $|{\bf T}|$, expressed in units of $eV_b/A$ (with $A$ denoting the area of a single hexagon in the graphene lattice), are shown in Fig.~\ref{fig:SOT} as functions of the magnetization direction ${\bf m}=(\sin\theta\cos\varphi,\sin\theta\sin\varphi,\cos\theta)$.

%%%%%%%%%%%%%%%%%%%%%%%%%%%%%%%%%%%%%%%%%%%%%%%%%%%%%%%%%%%%%%%%%%%%%%%%%%%%%%%%

\begin{table*}[t]
    \centering
    \setlength{\tabcolsep}{10pt}
    \begin{tabular}{l|ccccccccc}
     & $A_\nu^{(1)}$ & $\alpha_\nu^{(1)}$ & $A_\nu^{(2)}$ & $\alpha_\nu^{(2)}$ & $A_\nu^{(3)}$ & $\alpha_\nu^{(3)}$ & $A_\nu^{(4)}$ & $\alpha_\nu^{(4)}$ & $T_\nu^{(0)}$\\
     \hline\hline
     $T_x(\alpha,\varphi = 0)$ &  14.48 & 1.626 & 0 & 0 & 0 & 0 & 0 & 0 & 0 \\
     $T_y(\alpha,\varphi = 0)$ &  0.034 & 2.11 & -0.004 & -3.76 & -0.0008 & 1.98 & 0.0022 & 1.05 & 0.0026 \\
     $T_z(\alpha,\varphi = 0)$ &  -15.25 & 0 & 0 & 0 & 0 & 0 & 0 & 0 & 0 \\
     \hline
     $T_x(\alpha,\varphi = \pi/2)$ &  -14.93 & $\approx\pi/2$ & 0 & 0 & 0 & 0 & 0 & 0 & 0 \\
     $T_y(\alpha,\varphi = \pi/2)$ &  0.076 & 0.39 & 0.0094 & -3.87 & 0.0025 & 6.51 & 0.0005 & -0.93 & -0.004 \\
     $T_z(\alpha,\varphi = \pi/2)$ &  -0.096 & 1.29 & -0.012 & 0.72 & 0.004 & 1.54 & 0.001 & -6.12 & 0.0014 \\
     \hline
     $T_x(\theta = \pi/2,\alpha)$ &  -0.195 & 1.64 & -0.015 & 0.62 & -0.0019 & 2.30 & -0.0033 & 1.36 & 0.0005 \\
     $T_y(\theta = \pi/2,\alpha)$ &  0.065 & -0.25 & 0.0097 & 2.04 & -0.005 & 0.06 & -0.001 & -3.43 & -0.0025 \\
     $T_z(\theta = \pi/2,\alpha)$ &  -15.3 & $\approx\pi/2$ & 0 & 0 & 0 & 0 & 0 & 0 & 0 \\
     \hline
    \end{tabular}
    \caption{The relevant parameters from the fit, $A_\nu^{(n)}$ coefficients in units of $10^{-9}\,eV_b/A$, and $\alpha_\nu^{(n)}$ in radians, obtained by fitting the expansion of $\mathbf{T}$ in Eq.~(\ref{eq:series}) to the NEGF+ncDFT-computed angular dependence of SOT components.}
    \label{tab:coeffs}
\end{table*}

Assuming that current flows along the $x$-axis as in Fig.~\ref{fig:Device}, the nonzero torque can be expressed considering symmetry constraints as a series of powers $(\mathbf{e}_z\times\mathbf{m})$ \cite{garello2013symmetry}, where $\mathbf{e}_z$ is the unit vector along the $z$-direction. 
However, we observed that the angular dependences of the torque components are more complex, requiring a more general approach.
Here we fit the $\mathbf{T}$ components computed from NEGF+ncDFT formalism with a general function defined 
\begin{equation}\label{eq:series}
    T_\nu(\alpha) = T_\nu^{(0)} + \sum_n A_\nu^{(n)}\sin(n\alpha + \alpha_\nu^{(n)}),
\end{equation}
where $T_\nu^{(0)}$ is the torque offset, $\alpha_\nu^{(n)}$ are the phase shifts of the angular dependences, and $\nu=\{x,y,z\}$.
The relevant parameters fitted to the calculated SOT are given in Table~\ref{tab:coeffs}.
For the magnetization in the $xz$-plane, the torque is well approximated with the field-like dependence $(\mathbf{m}\times\mathbf{e}_y)$. 
We note that the $x$-component proportional to $\sim\sin(\theta+\theta_x^{(1)})$ of the torque has a global phase shift $\theta_x^{(1)} \approx 93.2^\circ$ related to the non-collinear direction between the magnetization in CrSBr and proximity-induced magnetic moment in graphene. 
The $y$-component has a more complex angular variation originating from the exchange and spin-orbit coupling proximity-induced field in graphene, and can associate properties of both field-like and damping-like torque~\cite{belashchenko2020interfacial}.
In the case of $yz$-plane magnetization variation, the $x$-component of the torque has the damping-like variation $\mathbf{m}\times(\mathbf{m}\times\mathbf{e}_y) \sim \cos(\theta)$. In contrast, the other components have significantly higher order contributions.
Similarly, for the $xy$-plane magnetization variation, the $x$ and $y$ components of the torque have complex angular variation, and the $z$-component is a field-like torque depending on $\mathbf{m}\times\mathbf{e}_y \sim \cos(\varphi)$.

\begin{figure}[b]
	\centering
	\includegraphics[width=1\columnwidth]{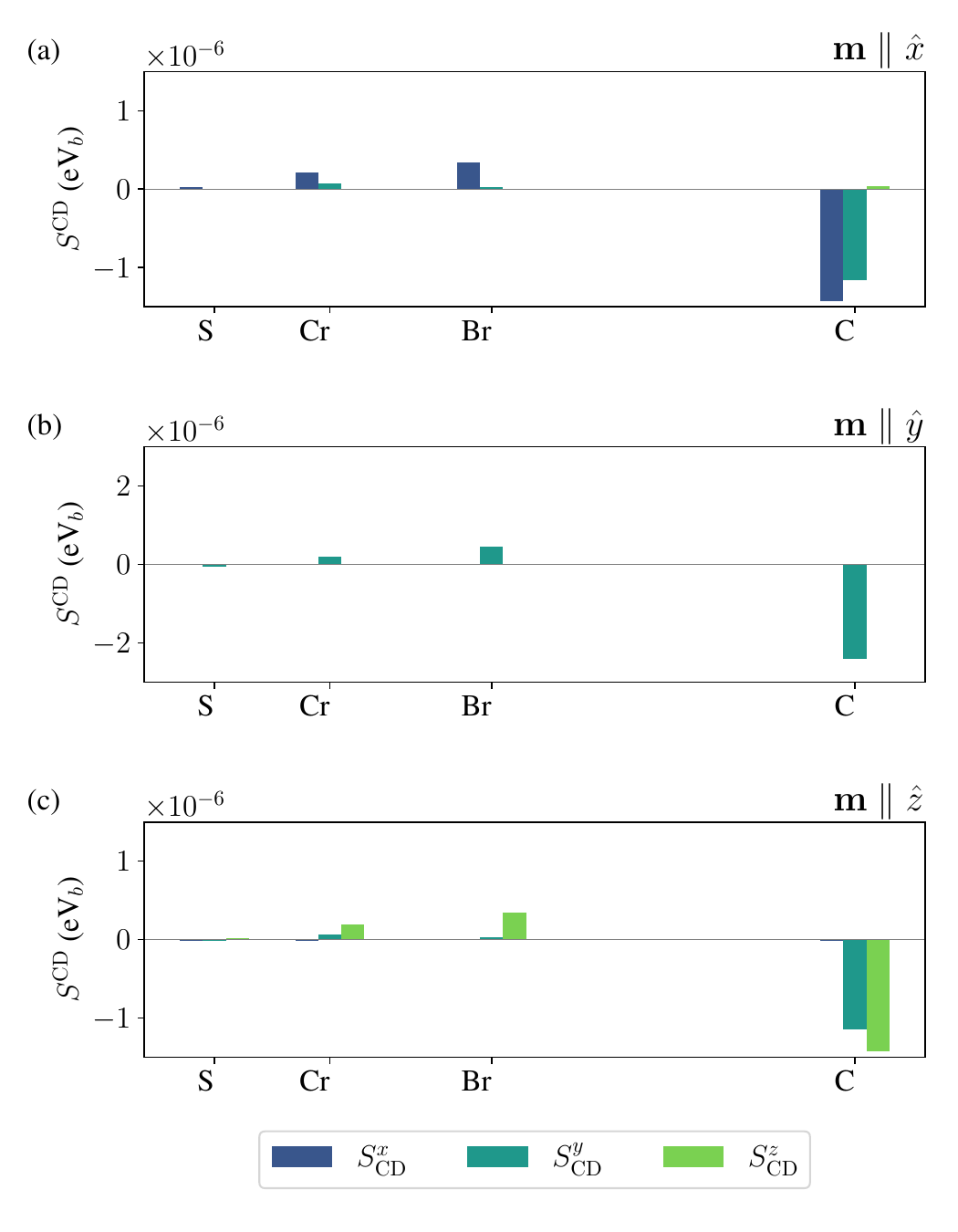}
	\caption{Calculated current-driven nonequilibrium spin density 
		$\mathbf{S}_{\mathrm{CD}} = (S_{\mathrm{CD}}^x, S_{\mathrm{CD}}^y, S_{\mathrm{CD}}^z)$ in the 
		linear-response regime in graphene on CrSBr monolayer
		for: (a) $\mathbf{m}\parallel \hat{x}$; (b) $\mathbf{m}\parallel \hat{y}$; and (c) $\mathbf{m}\parallel \hat{z}$.}
	\label{fig:S_CD}
\end{figure}

The current-driven nonequilibrium spin density $\mathbf{S}_{\mathrm{CD}}$ is plotted in Fig.~\ref{fig:S_CD} for three representative orientations
of the magnetization  $\mathbf{m}\parallel \{\hat{x}, \hat{y}, \hat{z}\}$. 
Under bias, the current flowing through the graphene gives rise to the development of the nonequilibrium spin density in graphene due to proximity-induced exchange and spin-orbit coupling fields. 
As the torque on the Dirac electrons is evaluated from that spin density crossed with the proximity-induced exchange field, the effect can be regarded as a proximity-induced self–spin–orbit torque. The self-induced SOT was reported recently in conventional devices \cite{Tang2020:AdvMat,Berrocal2021:AdvMat,Aoki2022:PRB,Hibino2024:PRB} with current flowing directly through the ferromagnetic layer or in a graphene-based van der Waals heterostructure \cite{zpd+20}.

\begin{figure}[t]
	\centering
	\includegraphics[width=1\columnwidth]{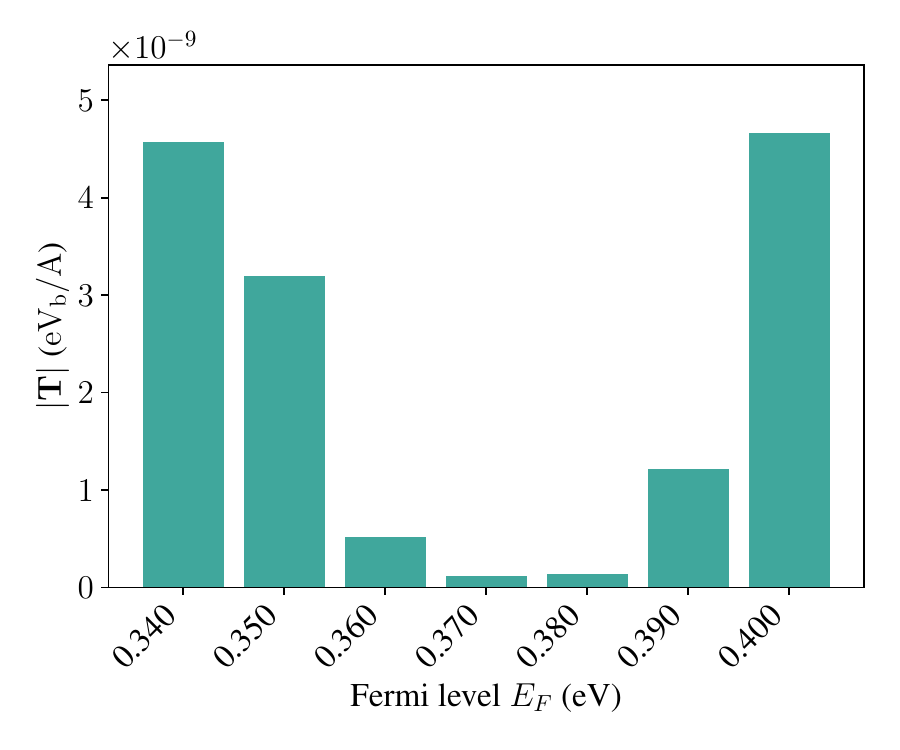}
	\caption{Total magnitude of the spin-orbit torque per unit hexagon area, $|\mathbf{T}|/A$, as a function of the Fermi level $E_F$ for the $\theta = 90^\circ$, $\varphi = 0^\circ$ magnetization orientation. The midpoint of the small gap opened at the Dirac point is located around $E \approx 0.375$~eV.}
	\label{fig:T_vs_EF}
\end{figure}

In addition to the angular dependence, we also analyze how the magnitude of the odd SOT varies with the Fermi level. Figure~\ref{fig:T_vs_EF} shows $|\mathbf{T}|/A$ as a function of $E_F$ for the representative orientation $\theta = 90^\circ, \varphi = 0^\circ$. The torque exhibits a pronounced sensitivity to the position of $E_F$. The midpoint of this gap lies at approximately $E \approx 0.375$~eV, which serves as a natural reference energy for comparing the torque response. 

\section{Conclusion}\label{conclusion}
In summary, we have performed a theoretical analysis of proximity-induced SOT in graphene/CrSBr heterostructure. 
Our DFT calculations show that the trigonal CrSBr monolayer is a ferromagnetic semiconductor with a band gap of about 1.1 eV and a Curie temperature of about 304~K, ensuring stable magnetism under technologically relevant conditions. 
When proximitized with graphene, the Dirac states spin split of about 20~meV due to exchange and spin-orbit coupling proximity-induced effects.
Quantum transport simulations reveal that the proximity-induced self-SOT in graphene is almost entirely governed by the odd component, reflecting the absence of spin Hall–driven even torques in this two-dimensional system. 
The phase shifts of the angular dependence of the SOT arise from the noncollinearity between the CrSBr magnetization and the induced magnetic moments in graphene.
Taken together, these findings identify graphene/CrSBr as a promising platform for realizing efficient, room-temperature spintronics devices.

%%%%%%%%%%%%%%%%%%%%%%%%%%%%%%%%%%%%%%%%%%%%%%%%%%%%%%%%%%%%%%%%%%%%%%%%%%%%%%%%
\vspace{0.5cm}
\section{Acknowledgment}
Funded by the European Union’s NextGenerationEU program through the Recovery and Resilience Plan for Slovakia under the Project No. 09I03-03-V04-00318.
%%%%%%%%%%%
M.R. acknowledges Red Española de Supercomputación for the computational resources provided by Universitat de València through the Project No. FI-2025-2-0019.
%%%%%%%%%%%
M. G. acknowledges funding by the EU NextGenerationEU through the Recovery and Resilience Plan for Slovakia under the project No. 09I05-03-V02-00071, and the Ministry of Education, Research, Development and Youth of the Slovak Republic, provided under Grant No. VEGA 1/0104/25, and the Slovak Academy of Sciences project IMPULZ IM-2021-42.

%%%%%%%%%%%%%%%%%%%%%%%%%%%%%%%%%%%%%%%%%%%%%%%%%%%%%%%%%%%%%%%%%%%%%%%%%%%%%%%%


\end{document}